\documentclass[aps,twocolumn,amsmath,superscriptaddress,amsfonts,amssymb,hyperref]{revtex4}
\usepackage[english]{babel}
\usepackage{graphicx}
\usepackage{dcolumn}
\usepackage{bm}
\usepackage{amssymb}
\usepackage{amsmath}
\usepackage{wasysym}
\usepackage{color}
\usepackage{times}
\usepackage{placeins}
\usepackage{epstopdf}

\newcommand{\dl}[2]	{$d^{#1}\underline{L}^{#2}$}			
\newcommand{\Jij}      {$J_{ij}$}

\newcommand{\ttg}      {$t_{2g}$}
\newcommand{\eg}       {$e_{g}$}
\newcommand{\ledg}     {$L_{2,3}$}

\newcommand{\lthr}     {$L_{3}$}
\newcommand{\medg}     {$M_{4,5}$}

\newcommand{\BiSeVII}  {V$_{0.1}$(Bi$_{0.32}$Sb$_{0.68}$)$_{1.9}$Te$_3$}

\newcommand{\BiSeC}    {Cr$_{0.1}$(Bi$_{0.1}$Sb$_{0.9}$)$_{1.9}$Te$_3$}
\newcommand{\vbst}     {V:(Bi,Sb)$_2$Te$_3$}
\newcommand{\cbst}     {Cr:(Bi,Sb)$_2$Te$_3$}
\newcommand{\vbstx}    {V$_z$(Bi$_{x}$Sb$_{1-x}$)$_{2-z}$Te$_3$}
\newcommand{\cbstx}    {Cr$_z$(Bi$_{x}$Sb$_{1-x}$)$_{2-z}$Te$_3$}
\newcommand{\vbt}   	 {V$_{0.1}$Bi$_{1.9}$Te$_3$}

\newcommand{\bst}   	 {(Bi,Sb)$_2$Te$_3$}
\newcommand{\bsxt}   	 {(Bi$_{x}$Sb$_{1-x}$)$_2$Te$_3$}
\newcommand{\bite}  	 {Bi$_2$Te$_3$}

\newcommand{\sbte}  	 {Sb$_2$Te$_3$}

\usepackage[normalem]{ulem}

\makeatletter
\renewcommand*\env@matrix[1][c]{\hskip -\arraycolsep
  \let\@ifnextchar\new@ifnextchar
  \array{*\c@MaxMatrixCols #1}}
\makeatother
\begin{document}

\title{Ubiquitous impact of localised impurity states on the exchange coupling mechanism in magnetic topological insulators}
\author{Thiago R. F. Peixoto\footnote{Corresponding author: Thiago.Peixoto@physik.uni-wuerzburg.de}}\affiliation{Experimentelle Physik VII and W\"urzburg-Dresden Cluster of Excellence ct.qmat, Fakult\"at f\"ur Physik und Astronomie, Universit\"at W\"urzburg, Am Hubland, D-97074 W\"urzburg, Germany}
\author{Hendrik Bentmann}\affiliation{Experimentelle Physik VII and W\"urzburg-Dresden Cluster of Excellence ct.qmat, Fakult\"at f\"ur Physik und Astronomie, Universit\"at W\"urzburg, Am Hubland, D-97074 W\"urzburg, Germany}
\author{Philipp R\"u{\ss}mann}\affiliation{Peter Gr\"unberg Institut (PGI-1) and Institute for Advanced Simulation (IAS-1), Forschungszentrum J\"ulich and JARA, D-52425 J\"ulich, Germany}
\author{Abdul-Vakhab Tcakaev}\affiliation{Experimentelle Physik IV and R\"ontgen Research Center for Complex Materials (RCCM), Fakult\"at f\"ur Physik und Astronomie, Universit\"at W\"urzburg, Am Hubland, D-97074 W\"urzburg, Germany}
\author{Martin Winnerlein}\affiliation{Experimentelle Physik III and Institut f\"ur Topologische Isolatoren, Fakult\"at f\"ur Physik und Astronomie, Universit\"at W\"urzburg, Am Hubland, D-97074 W\"urzburg, Germany}
\author{Steffen Schreyeck}\affiliation{Experimentelle Physik III and Institut f\"ur Topologische Isolatoren, Fakult\"at f\"ur Physik und Astronomie, Universit\"at W\"urzburg, Am Hubland, D-97074 W\"urzburg, Germany}
\author{Sonja Schatz}\affiliation{Experimentelle Physik VII and W\"urzburg-Dresden Cluster of Excellence ct.qmat, Fakult\"at f\"ur Physik und Astronomie, Universit\"at W\"urzburg, Am Hubland, D-97074 W\"urzburg, Germany}
\author{Raphael Crespo Vidal}\affiliation{Experimentelle Physik VII and W\"urzburg-Dresden Cluster of Excellence ct.qmat, Fakult\"at f\"ur Physik und Astronomie, Universit\"at W\"urzburg, Am Hubland, D-97074 W\"urzburg, Germany}
\author{Fabian Stier}\affiliation{Experimentelle Physik IV and R\"ontgen Research Center for Complex Materials (RCCM), Fakult\"at f\"ur Physik und Astronomie, Universit\"at W\"urzburg, Am Hubland, D-97074 W\"urzburg, Germany}
\author{Volodymyr Zabolotnyy}\affiliation{Experimentelle Physik IV and R\"ontgen Research Center for Complex Materials (RCCM), Fakult\"at f\"ur Physik und Astronomie, Universit\"at W\"urzburg, Am Hubland, D-97074 W\"urzburg, Germany}
\author{Robert J. Green}\affiliation{Department of Physics and Astronomy and Stewart Blusson Quantum Matter Institute, University of British Columbia, Vancouver, BC V6T 1Z4, Canada}\affiliation{Department of Physics and Engineering Physics and Centre for Quantum Topology and Its Applications (quanTA), University of Saskatchewan, SK S7N 5E2 Saskatoon, Canada}
\author{Chul Hee Min}\affiliation{Experimentelle Physik VII and W\"urzburg-Dresden Cluster of Excellence ct.qmat, Fakult\"at f\"ur Physik und Astronomie, Universit\"at W\"urzburg, Am Hubland, D-97074 W\"urzburg, Germany} 
\author{Celso I. Fornari}\affiliation{Experimentelle Physik VII and W\"urzburg-Dresden Cluster of Excellence ct.qmat, Fakult\"at f\"ur Physik und Astronomie, Universit\"at W\"urzburg, Am Hubland, D-97074 W\"urzburg, Germany}
\author{Henriette Maa\ss}\affiliation{Experimentelle Physik VII and W\"urzburg-Dresden Cluster of Excellence ct.qmat, Fakult\"at f\"ur Physik und Astronomie, Universit\"at W\"urzburg, Am Hubland, D-97074 W\"urzburg, Germany}
\author{Hari Babu Vasili}\affiliation{ALBA Synchrotron Light Source, E-08290 Cerdanyola del Vall\`{e}s, Spain}
\author{Pierluigi Gargiani}\affiliation{ALBA Synchrotron Light Source, E-08290 Cerdanyola del Vall\`{e}s, Spain}
\author{Manuel Valvidares}\affiliation{ALBA Synchrotron Light Source, E-08290 Cerdanyola del Vall\`{e}s, Spain}
\author{Alessandro Barla}\affiliation{Istituto di Struttura della Materia (ISM), Consiglio Nazionale delle Ricerche (CNR), I-34149 Trieste, Italy}
\author{Jens Buck}\affiliation{Deutsches Elektronen-Synchrotron DESY, Hamburg, Germany}
\author{Moritz Hoesch}\affiliation{Deutsches Elektronen-Synchrotron DESY, Hamburg, Germany}
\author{Florian Diekmann}\affiliation{Institut f\"ur Experimentelle und Angewandte Physik, Christian-Albrechts-Universit\"at zu Kiel, 24098 Kiel, Germany}
\author{Sebastian Rohlf}\affiliation{Institut f\"ur Experimentelle und Angewandte Physik, Christian-Albrechts-Universit\"at zu Kiel, 24098 Kiel, Germany}
\author{Matthias Kall\"ane}\affiliation{Institut f\"ur Experimentelle und Angewandte Physik, Christian-Albrechts-Universit\"at zu Kiel, 24098 Kiel, Germany}
\affiliation{Ruprecht Haensel Laboratory, Kiel University and DESY, Germany}
\author{Kai Rossnagel}\affiliation{Deutsches Elektronen-Synchrotron DESY, Hamburg, Germany}\affiliation{Institut f\"ur Experimentelle und Angewandte Physik, Christian-Albrechts-Universit\"at zu Kiel, 24098 Kiel, Germany}\affiliation{Ruprecht Haensel Laboratory, Kiel University and DESY, Germany} 
\author{Charles Gould}\affiliation{Experimentelle Physik III and Institut f\"ur Topologische Isolatoren, Fakult\"at f\"ur Physik und Astronomie, Universit\"at W\"urzburg, Am Hubland, D-97074 W\"urzburg, Germany}
\author{Karl Brunner}\affiliation{Experimentelle Physik III and Institut f\"ur Topologische Isolatoren, Fakult\"at f\"ur Physik und Astronomie, Universit\"at W\"urzburg, Am Hubland, D-97074 W\"urzburg, Germany}
\author{Stefan Bl\"ugel}\affiliation{Peter Gr\"unberg Institut (PGI-1) and Institute for Advanced Simulation (IAS-1), Forschungszentrum J\"ulich and JARA, D-52425 J\"ulich, Germany}
\author{Vladimir Hinkov}\affiliation{Experimentelle Physik IV and R\"ontgen Research Center for Complex Materials (RCCM), Fakult\"at f\"ur Physik und Astronomie, Universit\"at W\"urzburg, Am Hubland, D-97074 W\"urzburg, Germany}
\author{Laurens W. Molenkamp}\affiliation{Experimentelle Physik III and Institut f\"ur Topologische Isolatoren, Fakult\"at f\"ur Physik und Astronomie, Universit\"at W\"urzburg, Am Hubland, D-97074 W\"urzburg, Germany}
\author{Friedrich Reinert}\affiliation{Experimentelle Physik VII and W\"urzburg-Dresden Cluster of Excellence ct.qmat, Fakult\"at f\"ur Physik und Astronomie, Universit\"at W\"urzburg, Am Hubland, D-97074 W\"urzburg, Germany}

\date{December 16,2019}

\maketitle
\textbf{Since the discovery of the quantum anomalous Hall effect in the magnetically doped topological insulators (MTI) \cbst~\cite{chang:13} and \vbst~\cite{chang:15}, the search for the exchange coupling mechanisms underlying the onset of ferromagnetism has been a central issue, and a variety of different scenarios have been put forward \cite{yu:10,Zhang2012,Vergniory2014,chang:13,chang:15,Li:15,Kou2013,Kou2015,Kim:17,Kim2018,ye:15,Ye2019,Ruessmann2018}. By combining resonant photoemission, X-ray magnetic dichroism and multiplet ligand field theory, we determine the local electronic and magnetic configurations of V and Cr impurities in \bst. While strong $pd$ hybridisation is found for both dopant types, their $3d$ densities of states show pronounced differences. %, forming a narrow peak near the Fermi level for V and a broad distribution over the host valence band for Cr. 
State-of-the-art first-principles calculations show how these impurity states mediate characteristic short-range $pd$ exchange interactions, whose strength sensitively varies with the position of the $3d$ states relative to the Fermi level. Measurements on films with varying host stoichiometry support this trend. Our results establish the essential role of impurity-state mediated exchange interactions in the magnetism of MTI. 
} 

Magnetically doped topological insulators (MTI) form a cornerstone in the field of topological quantum materials. Particularly in Cr- and V-doped \bst, the combination of ferromagnetism and a topologically non-trivial electronic band structure led to the discovery of the quantum anomalous Hall (QAH) effect \cite{chang:13,Kou2015,chang:15,Bestwick:15,Grauer:15}, \textit{i.e.} a dissipationless quantised edge-state transport in the absence of external magnetic fields. These materials are now being widely utilized for possible realisations of topological superconductor \cite{Tokura2019} and axion insulator states \cite{Xiao2018}, as well as in the context of metrology \cite{Goetz2018} and spintronic functionalities \cite{Fan2014}. However, despite the broad interest in these dilute MTI, controversy still remains as to the microscopic origin of the ferromagnetism and to the electronic states inducing the ferromagnetic (FM) coupling.

In a pioneering work, predicting the QAH effect in MTI, the FM state was proposed to arise from a van Vleck mechanism, as a result of strong spin-orbit coupling (SOC) and the topologically non-trivial band ordering in these materials \cite{yu:10}. Although some experimental support for this scenario has been reported \cite{Li:15,chang:13_2,Kou2015}, more recent first-principles calculations find that the strength of the exchange interactions in \cbst~and \vbst~is, in fact, largely independent of SOC, suggesting that the van Vleck mechanism, in the form proposed in \cite{yu:10}, plays no decisive role \cite{Kim:17,Kim2018}. Other theoretical works predict a dependence of the magnetic coupling on the precise configuration of the impurity 3$d$ states \cite{Zhang2012,Zhang2013,Vergniory2014,Kim:17,Kim2018}, as it is known in the context of dilute magnetic semiconductors \cite{Sato2010}. Experimentally the robustness of the FM state in \cbst~and \vbst~films was indeed found to vary substantially with dopant type and host stoichiometry \cite{chang:15,ye:15,Winnerlein2017,Ye2019}. So far, however, a comprehensive understanding of how this behaviour is related to the local impurity electronic structure is still absent. 

In this work, we present a combination of key experiments and theory that establishes the fundamental link between local impurity electronic structure and magnetic coupling in MTI. 
By means of X-ray magnetic circular dichroism (XMCD) and resonant photoelectron spectroscopy (resPES) we systematically probe the electronic and magnetic fingerprints of the $3d$ states of V and Cr impurities embedded in the bulk of \bsxt~thin films, as well as the effect of Bi/Sb substitution in the host. Supported by multiplet ligand field theory (MLFT) and \textit{ab initio} density functional theory (DFT) calculations, our results unveil the essential role of impurity-state-mediated exchange interactions underlying the magnetic properties of V- and Cr-doped \bst~thin films. 
%
%%------Fig. 1------%%
\begin{figure*}[htb]
\centering
\includegraphics[width=6.9in]{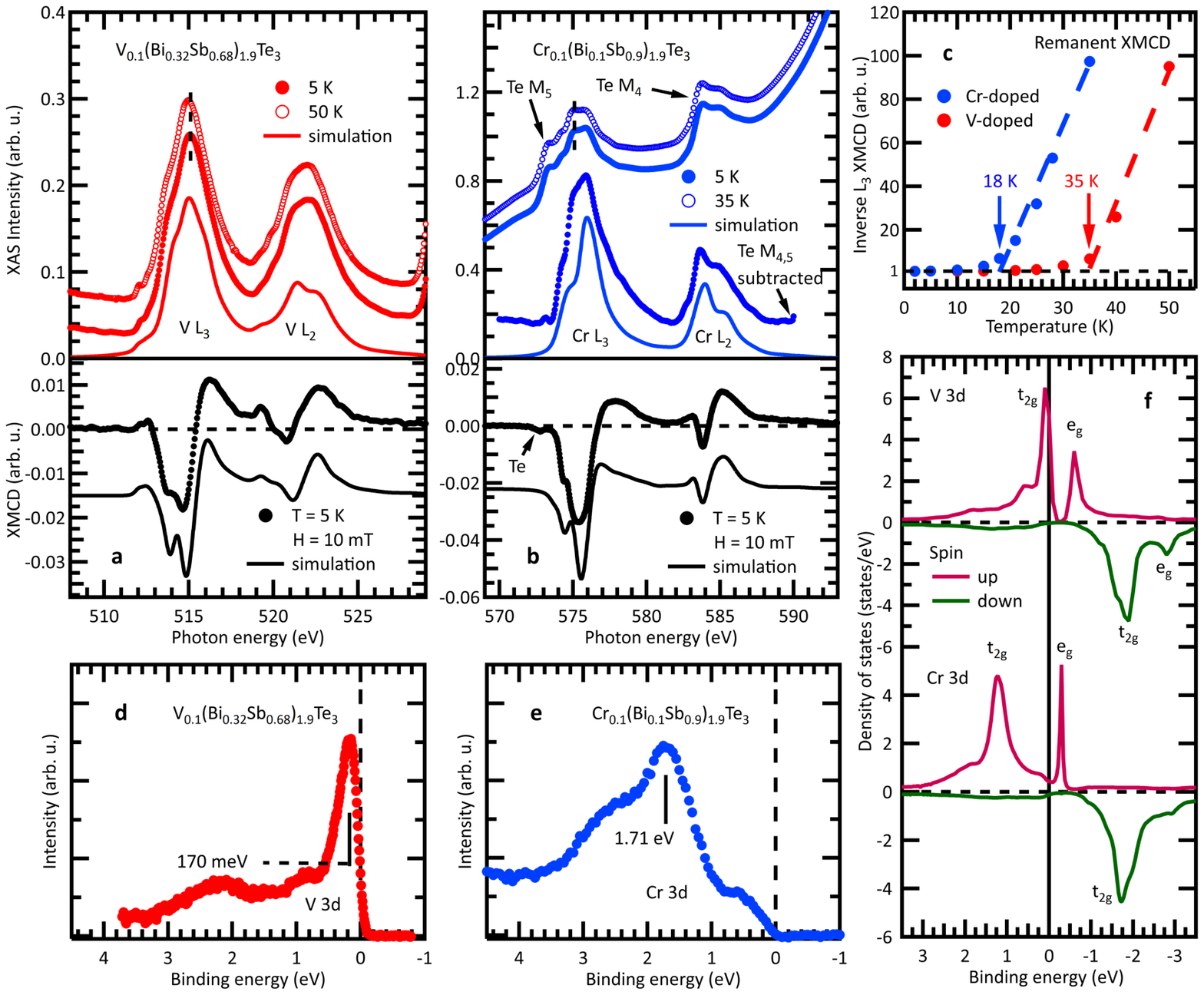}
\caption{\label{fig1}\textbf{Electronic and magnetic fingerprints of V and Cr impurities in \bst.} XAS (upper panels) and XMCD (lower panels) spectra of \textbf{a,} \BiSeVII~and \textbf{b,} \BiSeC~thin films, respectively at the V and Cr \ledg~edges, taken at temperatures above (hollow circles) and below (full circles) $T_C$. 
The contribution of the Te \medg~edges to the Cr \ledg~spectrum, pointed out by the arrows in \textbf{b}, was subtracted according to a reference spectrum measured on an undoped sample. The XMCD spectra (black circles) were measured in the remanent state, at 5 K, demonstrating a FM state for both systems. 
The corresponding continuous curves are spectra calculated by multiplet ligand field theory. 
The small dip in the pre-edge of the Cr \lthr~XMCD is attributed to the induced magnetic moment in the Te atoms. 
The model found a $d$-shell occupation of $d^{3.13}$ ($m_S=2.33$ $\mu_B$/atom) for V and $d^{4.37}$ ($m_S=3.32$ $\mu_B$/atom) for Cr.
\textbf{c,} Temperature-dependence of the inverse remanent XMCD signal at the V (red circles) and Cr (blue circles) \lthr~edges, from which $T_C(\text{V})\sim 35$ K and $T_C(\text{Cr})\sim 18$ K are estimated. 
resPES spectra at the \lthr~edges of \textbf{d,} V ($h\nu=515$ eV) and \textbf{e,} Cr ($h\nu=575.6$ eV), at 30 K, representing the occupied $3d$ DOS. 
\textbf{f,} Calculated spin-up (magenta curves) and spin-down (green curves) $3d$ DOS for substitutional V and Cr single impurities in \sbte. The exchange-split \ttg~and \eg~manifolds are identified. 
}
\end{figure*} 

We start by presenting in Figs. \ref{fig1}a and b the X-ray absorption (XAS) and XMCD spectra from \BiSeVII~and \BiSeC, at the V and Cr \ledg~edges, measured at temperatures above (hollow circles) and below (full circles) the Curie temperature $T_C$ (cf. Fig. \ref{fig1}c). No energy shifts or changes in the branching ratio are observed at the \ledg~edges with varying temperature across $T_C$, contradicting recent reports where a putative energy shift was interpreted as evidence of the van Vleck mechanism \cite{Li:15}. The XMCD spectra were measured under a small applied magnetic field of 10 mT (remanent state), oriented perpendicular to the surface, at 5 K. They confirm a persistent FM state at this temperature and a sizable magnetic moment carried by the transition metal (TM) $3d$ states for both V and Cr doped \bst~systems. 
From the temperature-dependent XMCD signal in Fig. \ref{fig1}c, we deduce roughly a factor of 2 difference in $T_C$ for V and Cr at similar doping concentrations, in good agreement with previous works \cite{chang:15,ye:15}. 

We performed multiplet ligand field theory calculations to analyse the V and Cr \ledg~lineshapes, assuming substitutional Cr and V atoms in an octahedral crystal-field. 
As seen in Figs. \ref{fig1}a,b, the calculations reproduce all main features of the measured spectra. 
We find a strong deviation from the 3+ ionic (V $d^2$, Cr $d^3$) ground state, with about 90\% weight of charge transfer states with one and two holes in the ligand orbitals. 
The resulting $d$-shell occupation is $d^{3.13}$ ($m_S=2.33$ $\mu_B$/atom) for V and $d^{4.37}$ ($m_S=3.32$ $\mu_B$/atom) for Cr. Correspondingly, our first-principles calculations for single V and Cr impurities embedded in a quintuple layer of \sbte, on an octahedral Sb site, find $d^{3.3}$ ($m_S\approx2.55~\mu_B$) and $d^{4.4}$ ($m_S\approx3.75~\mu_B$). 
This implies a substantial charge transfer from the ligand $p$ states into the $3d$ impurity states as a result of $pd$ hybridisation. 

We next assess the contribution of the $3d$ states to the valence band (VB) by resPES at the V and Cr \lthr~edges (Figs. \ref{fig1}d,e). The data sets are obtained from the normalized difference between the on- and off-resonant photoemission spectra, taken at 30 K, and represent the V and Cr $3d$ partial density of states (DOS). The data establish a remarkably pronounced difference in the character of the $3d$ DOS for V and Cr impurities. For V, the $3d$ states pile up predominantly in a narrow peak just below $E_F$ with the center at $E_B=170$ meV. For Cr, on the other hand, the $3d$ states are broadly distributed over the host VB, with a maximum at $E_B=1.71$ eV and low spectral weight near $E_F$. Our first-principles calculations nicely capture these main characteristics. Fig. \ref{fig1}f shows the calculated spin-polarised $3d$ DOS for V (upper panel) and Cr (lower panel) impurities in \sbte, which allow us to assign the experimentally observed states to majority \ttg~states. Our findings yet support a recent scanning tunneling microscopy and spectroscopy study of the electronic fingerprints of V and Cr impurities in \sbte~\cite{Zhang2018}. 

The presence of exchange-split $3d$ states at $E_F$ suscitate the emergence of localised magnetic moments and, possibly, long-range magnetic order. We now present our calculations of the magnetic exchange interactions. To this end, we first consider the spin-up DOS of V and Cr impurities in \sbte, calculated for different $E_F$ shifts ($\Delta E_F$), \textit{i.e.} different positions of $E_F$ with respect to the bulk VB and conduction band (CB). This approach allows us to simulate the effect of $n$- and $p$-type charge doping on the $3d$ states, \textit{i.e.} the valence of the TM impurities \cite{Ruessmann2018}. We find that the position of the impurity states depends sensitively on the charge doping, shifting towards higher binding energies when going from $p$-type (dark coloured curves) to $n$-type (light coloured curves) doping. In particular, the narrow V $3d$ peak gradually moves away from $E_F$. This trend is supported experimentally by our resPES data in Fig. \ref{fig3}a for two \bsxt~films with different Bi concentrations $x$. For higher $x$, and thus higher $n$-doping \cite{Zhang2012,Zhang2013,chang:15}, the V $3d$ peak shifts to significantly higher binding energies. 

\begin{figure*}[htb]
\centering
\includegraphics[width=7.0in]{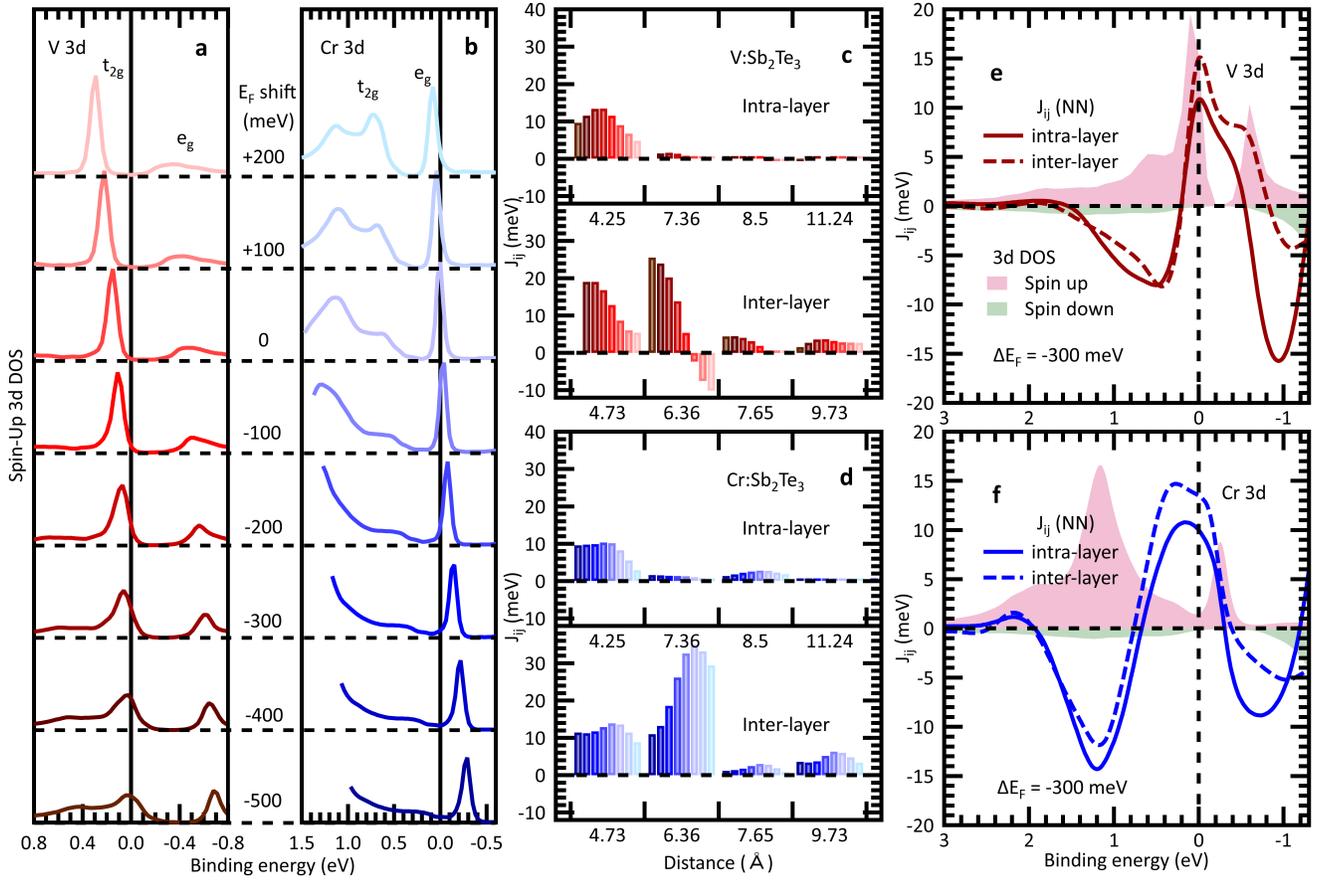}
\caption{\label{fig2}\textbf{First-principles calculations of magnetic interactions in V- and Cr-doped \sbte.} Spin-up partial $3d$ DOS of \textbf{a,} V and \textbf{b,} Cr impurities embedded in \sbte, calculated for different positions of $E_F$ in respect to the bulk VB and CB ($\Delta E_F$). Intra-layer (upper panels) and inter-layer (lower panels) exchange-coupling constants \Jij~as a function of distance from the \textbf{c,} V and \textbf{d,} Cr impurities, calculated for the same $\Delta E_F$ as in \textbf{a} and \textbf{b} (note the corresponding colour code). 
Energy-dependence of the intra-layer (solid line) and inter-layer (dashed line) NN \Jij$(E)$ for \textbf{e,} V:\sbte~and \textbf{f,} Cr:\sbte, at $\Delta E_F=-300$ meV. The corresponding spin-up (spin-down) $3d$ DOS is also plotted as light red (light green) shaded area. \Jij$(E)$ exhibits a predominant FM character around $E_F$, and consists of a sharp peak coinciding with the localised \ttg~peak for V doping (absent in the Cr case), and a single broad ridge extending between the onsets of the \ttg~and \eg~manifolds for Cr doping. 
These two distinct features are attributed to intrinsically different impurity-state-mediated $pd$ exchange interactions.}
\end{figure*}

\begin{figure}[htb]
\centering
\includegraphics[width=3.495in]{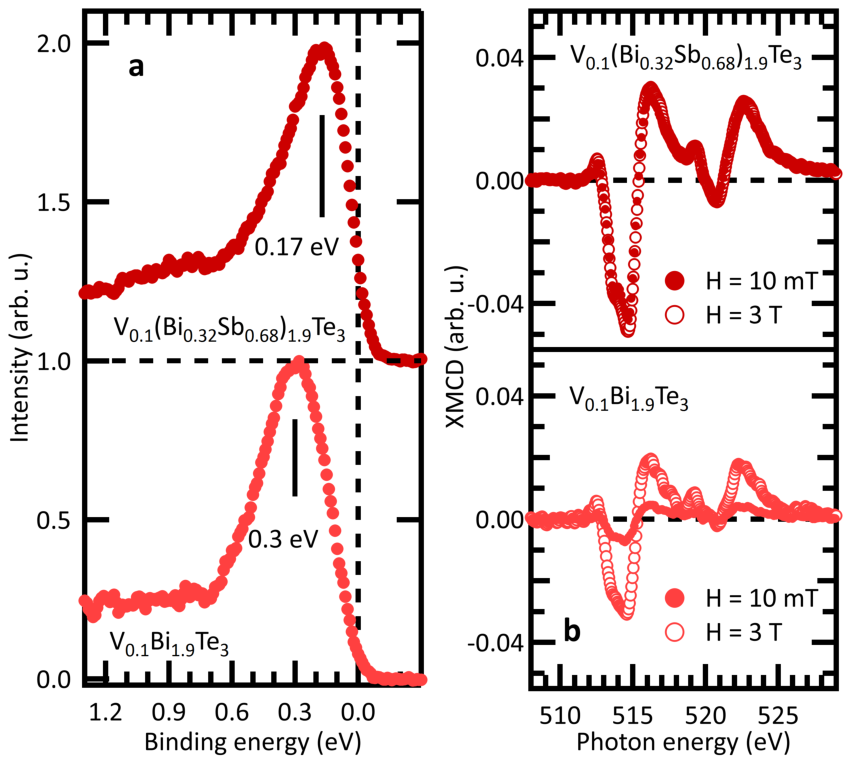}
\caption{\label{fig3}\textbf{Impact of the $3d$ impurity states on the magnetic properties of \vbst.} \textbf{a,} resPES spectra of the V $3d$ states from \BiSeVII~(dark red curve) and \vbt~(light red curve) thin films. The V $3d$ peak is shifted from $E_B=170$ meV in the former to $E_B=300$ meV in the latter. 
\textbf{b}, V \ledg~XMCD spectra from the same samples in \textbf{a}, taken at remanence (full circles) and saturation (hollow circles), at 5 K. The remanent XMCD signal remains high in the Sb-rich phase, whereas it is strongly suppressed in the Bi-rich one.   
}
\end{figure}

In Figs.~\ref{fig2}c,d we plot the calculated exchange-coupling constants \Jij~for different separation distances between two TM ions located in one atomic plane (intra-layer coupling, upper panels) and at neighbouring atomic planes within the same quintuple layer (inter-layer coupling, lower panels), using the same $\Delta E_F$ values as in Figs. \ref{fig2}a,b. 
For both V and Cr, the interlayer coupling is dominant, in good agreement with ref. \cite{Vergniory2014}. Overall, at nearest neighbour (NN) distances the calculated \Jij~contributions are larger for V, while at the second and third neighbour positions the \Jij~values for Cr are markedly larger. This behaviour follows the spatial decay of the calculated impurity-induced features in the $3d$ DOS. Most importantly, we find that \Jij~strongly varies with charge doping. 
Already small shifts of $E_F$ give rise to considerable changes in \Jij, whose sign and strength are linked to characteristic features in the $3d$ DOS at $E_F$. 
For V, \Jij~rapidly decreases in the $n$-doped regime, while for Cr the behaviour is more complex and depends more strongly on the relative impurity positions. 
Figs. \ref{fig2}e,f show the energy dependence of \Jij$(E)$ for NN spins for $\Delta E_F=-300$ meV, which best corresponds to our experimental resPES data. The inter-layer and intra-layer components are respectively depicted as dashed and solid lines, and they exhibit qualitatively similar trends. In the same plots, the corresponding spin-up and spin-down $3d$ DOS are shown as shaded areas. 
For V, \Jij$(E)$ near $E_F$ consists of a sharp peak overlapping with a broad flat ridge that spans between the onsets of the \ttg~and \eg~peaks. \Jij$(E)$ is maximized when $E_F$ is positioned inside the V \ttg~manifold, and rapidly decreases otherwise. For Cr, the sharp peak in \Jij$(E)$ is absent, and only the broad ridge is seen. In sharp contrast to V, we find for Cr a maximal \Jij$(E)$ when $E_F$ lies in the gap between the \ttg~and \eg~peaks, where the $3d$ DOS is low. These distinct features in the V and Cr $3d$ DOS are fully in line with our measurements in Fig. \ref{fig1}. In the context of dilute magnetic semiconductors, similar characteristics in the exchange coupling constants have been extensively discussed \cite{Larson1988,Kacman2001,Belhadji2007,Sato2010,Dietl2010}. By comparison, we may tentatively associate the sharp peak in \Jij$(E)$ for V with the double-exchange mechanism and the broad ridge, found for both dopant types, with the FM superexchange mechanism \cite{Belhadji2007,Sato2010,Kim:17,Kim2018}.
%
%%------Fig. 2------%%

Our resPES and XMCD data from \BiSeVII~(dark red curve) and \vbt~(light red curve) in Fig. \ref{fig3} further demonstrate the strong sensitivity of the magnetic coupling on the position of the V impurity states. The V $3d$ maximum is found at $E_B=170$ meV in the former and at $E_B=300$ meV in the latter. This may be attributed to the effect of charge transfer from the Bi ions into the V impurities \cite{Zhang2011,Zhang2013,Kellner2015}. Fig. \ref{fig3}b shows a comparison of V \ledg~XMCD spectra for the same samples as in Fig. \ref{fig3}a, measured at saturation (hollow circles) and in remanence (full circles) at 5 K. While in Sb-rich \vbst~the remanent and saturated spectra almost coincide, in \vbt~the remanent XMCD is critically lower than in saturation, demonstrating a suppression of ferromagnetic interactions coinciding with the shift of the V states away from $E_F$, as predicted in our theoretical calculations. The observed weakening of the ferromagnetism upon Bi doping in \vbst~is in agreement with previous works \cite{chang:15,Winnerlein2017,Ye2019}, as well reported in single-crystalline \cbst~\cite{ye:15}. Our calculations in Figs. \ref{fig2}d,f also indicate a less pronounced dependence of the magnetic coupling on the $E_F$ position for Cr. This finding may explain the stronger gate-voltage dependence of the magnetic properties observed in V-doped \cite{chang:15} as compared to Cr-doped films \cite{chang:13,chang:13_2}.
%
%%------Fig. 3------%%

Our systematic experimental and theoretical results highlight the central role of impurity-state mediated exchange coupling for the magnetism in the paradigmatic QAH insulators \cbst~and \vbst. 
The latter cannot be explained based solely on the van Vleck mechanism, bearing its origin on the topologically non-trivial band structure \cite{yu:10,chang:13_2,Li:15}. 
Instead, our theoretical calculations on the basis of $pd$ hybridisation and $pd$ exchange coupling unambiguously elucidate the experimental observations. They show that the nature and strength of the magnetic exchange coupling vary with the position of $E_F$ in the $3d$ DOS, \textit{i.e.} with the occupation of the $3d$ states, thereby reconciling, in a unified theory, the differences observed between V and Cr doping of \bsxt~films and the host stoichiometry dependence of the magnetic properties of QAH insulators. 

\section{Methods}
\textbf{Sample preparation, structural, transport and magnetic characterisation}. Thin films (about 9 to 10 nm thick) of \cbstx~and \vbstx~were grown by molecular beam epitaxy (MBE) on hydrogen passivated Si(111) substrates. The growth details and characterisation, \textit{e.g.} by X-ray diffraction, atomic force microscopy and Hall magnetotransport, confirming the realisation of QAH effect in the V-doped samples (for $x=0.76-0.79$ and $z=0.1-0.2$), are published elsewhere \cite{Grauer:15,Grauer:17,Tarakina2017,Winnerlein2017}. After growth, the films were capped by a protective Te layer ($\sim100$ nm), which was mechanically removed in UHV conditions, prior to the spectroscopic measurements. Recent results have demonstrated the effectiveness of this decapping method on \bite~layers with high pristine quality \cite{Fornari2016}. The stoichiometries applied in this work are comparable to those that exhibited a stable and reproducible QAH effect.  

\textbf{XAS, XMCD and resPES}. The XAS and XMCD data was acquired at the HECTOR endstation, located at BOREAS beamline of the ALBA storage ring (Barcelona, Spain) \cite{Barla2016}. The measurements were performed in total electron yield mode, under magnetic fields of up to 6 T and temperatures down to 5 K. The resPES experiments were conducted at the ASPHERE III endstation located at beamline P04 of the PETRA III storage ring of DESY (Hamburg, Germany). The on- and off-resonant VB photoemission spectra were taken at $h\nu_{on}=514.8$ eV and $h\nu_{off}=508$ eV for V doped samples, and at $h\nu_{on}=575.6$ eV and $h\nu_{off}=560$ eV for Cr doped ones, according to the respective XAS spectra in Fig. \ref{fig1}). The energy resolution of the resPES measurements was typically better than 67 meV. All experiments were performed in ultra-high vaccuum (UHV) at pressures below $3\times 10^{-10}$ mbar. 

\textbf{DFT calculations}. The \sbte~and \bite~bulk crystals were simulated using the experimental bulk lattice structure (see ref. \cite{Ullner1968} for \sbte~and ref. \cite{Nakajima1963} for \bite). The electronic structure was calculated within the local density approximation (LDA) \cite{Vosko1980} to DFT by employing the full-potential relativistic Korringa-Kohn-Rostoker Green's function method (KKR) \cite{Ebert2011,Bauer2013} with exact description of the atomic cells \cite{Stefanou1990,Stefanou1991}. The truncation error arising from an $\ell_{\mathrm{max}}= 3$ cutoff in the angular momentum expansion was corrected for using Lloyd's formula \cite{Zeller2004}.
The V and Cr defects, together with a charge-screening cluster comprising the first two shells of neighbouring atoms (consisting of about 20 surrounding scattering sites), were embedded self-consistently using the Dyson equation in the KKR method \cite{Bauer2013} and have been chosen to occupy the substitutional Sb/Bi position in the quintuple layers and structural relaxations were neglected, while keeping the direction of the impurity's magnetic moment fixed along the out-of-plane direction. The shift in the Fermi level occurring in \bst~was accounted for by adjusting the self-consistently computed Fermi level of the host systems in the impurity embedding step.
The exchange interactions among two impurities were computed using the method of infinitesimal rotations \cite{Liechtenstein1987}, which map the exchange interaction to the Heisenberg Hamiltonian $\mathcal{H}=-\tfrac{1}{2} \sum_{\left\langle ij \right\rangle} J_{ij} \vec{s}_i \cdot \vec{s}_j$.

\textbf{Multiplet ligand field theory calculations}. Theoretical XAS and XMCD spectra for the \ledg~$(2p \rightarrow 3d)$ absorption edges of V and Cr ions were calculated by means of a configuration interaction (CI) cluster model, considering the central TM ion surrounded by six ligands (Te anions). We take into account all the $2p-3d$ and $3d-3d$ electronic Coulomb interactions, as well as the SOC on every open shell of the absorbing atom. We consider nominal $2p^63d^n$ ($n=2$ for V$^{3+}$and $n=3$ for Cr$^{3+}$) configurations and further include three more charge transfer (CT) states \dl{n+1}{1}, \dl{n+2}{2}~and \dl{n+3}{3}~($\underline{L}^1$ denotes a hole in the Te $5p$ orbitals) to account for hybridisation effects. To perform the CI calculation, the following fit parameters were introduced: scaling parameter $\beta$ for the Hartree-Fock values of the Slater integrals, the CT energy $\Delta$, the Coulomb interaction energy $U_{dd}$ between the $3d$ electrons, the hybridisation energy $V_{eg}$ and the octahedral crystal field parameter $10Dq$.
The simulations were performed using the \textit{Quanty} software for quantum many-body calculations, developed by M. W. Haverkort \cite{Haverkort2012}.
We assume V/Cr ions embedded in the cation sites and describe the crystal field in $O_h$ symmetry, with $C_4$ axes of the octahedron along the V-Te bonds. 
The spectral contributions from each of the split ground state terms to the absorption spectra were weighted by a Boltzmann factor. The calculated spectra were broadened by a Gaussian function to account for the instrumental broadening and by an energy-dependent Lorentzian profile for intrinsic lifetime broadening.

\bibliographystyle{apsrev}

\section{Acknowledgments}
T.R.F.P. would like to thank Kai Fauth and Arthur Ernst for their collaboration at early phases of this study. H.B. would like to thank J\'an Min\'ar, and P. R. would like to thank Peter Dederichs for helpful discussions. 
P.R. and S.B. acknowledge funding from the Priority Programme SPP-1666 Topological Insulators of the Deutsche Forschungsgemeinschaft (DFG) (projects MA4637/3-1) and from the VITI Programme of the Helmholtz Association, as well as computing time granted by the JARA Vergabegremium and provided on the JARA Partition part of the supercomputer CLAIX at RWTH Aachen University.
R.J.G. was supported by the Natural Sciences and Engineering Research Council of Canada. 
We acknowledge the financial support from the DFG through the SFB1170 `ToCoTronics' and the Würzburg-Dresden Cluster of Excellence on Complexity and Topology in Quantum Matter – \textit{ct.qmat} (EXC 2147, project-id 39085490).

\end{document}